\documentclass{ws-p8-50x6-00}
\renewcommand{\rightnote}{}
\begin{document}

\title{An introduction to 
lattice QCD \\ at non--zero temperature 
and density}

\author{Maria-Paola Lombardo}

\address{Istituto Nazionale di Fisica Nucleare -- Sezione di Padova}  


\maketitle

\abstracts{This is an informal overview of methods and results
on the QCD phase diagram
and lattice termodynamics
aimed at specialists in nearby fields.}

\section{Introduction}

QCD at non-zero temperature and density is a rich
subject discussed in many talks at this meeting:
from the introductory lesson describing the theoretical
 discovery  of
the quark gluon plasma \cite{cabibbo}  to the subtle interrelations
of gauge fields dynamics and charmonium suppresions \cite{satz},
from the detailed analysis of the exotic phases at large baryon density
\cite{nardulli} to the fascinating aspects of the QCD phase diagram
with many flavor \cite{sannino}, from the time-honored theoretical 
subjects of the interplay  between topology and confinement
\cite{deldebbio} to the challenging aspects of the interrelations
of chiral and axial symmetries  \cite{meggiolaro}.
These  notes  mostly aimed at the non-specialist
might hopefully provide, besides some introductory material,
a path  among general idea as well as among the many
new results presented in the lively
topical session on the QCD phases.

The material is organised as follows: 
Section 2 reviews the imaginary time formulation of 
field theory thermodynamics, and gives
the functional integral representation
of the partition function $\cal Z$. 
The representation of $\cal Z$ 
naturally leads to a theoretical suggestion:
the concept of universality which is singled out in a small
Section on its own, because it is so important.
Section 4 discusses the QCD chiral transition, 
Section 5 the QCD deconfinement transition, and Section 6
how to put together the two, eventually describing the
complex phenomenology of the real world
high temperature QCD phase transition. The equation
of state is briefly mentioned here. Section 6 introduces the method
used to obtain all of the results of the previous
Sections: the lattice regularization. 
We shall see that introducing temperature
is  straightforward while a non--zero density
poses specific problems. We will only mention
the main proposals to circumvent such problems, referring to
the most recent reviews 
\cite{kogut} for details and recent results.

\section{Formulation}
Here we will concern ourselves with the path integral 
representation of the partition function.
The basic property of equilibrium field theory is that one
single function $\cal Z$ ( the grand canonical partition function):
\begin{equation}
\cal Z = \cal Z (V, T, \mu)
\end{equation}
determines completely
the thermodynamic state of a system  according to:
\begin{eqnarray}
P &=& T \frac {\partial ln {\cal Z}}{\partial V} \\
N &=& T \frac {\partial ln {\cal Z}}{\partial \mu} \\
S &=& \frac {\partial T ln {\cal Z}}{\partial T} \\
E &=& -PV + TS +\mu N 
\end{eqnarray}
while physical observables $<O>$ can be computed as
\begin{equation}
<O> = Tr O \hat \rho / {\cal Z}
\end{equation}

Any of the excellent books on statistical field theory and thermodynamics
can provide a more detailed discussion of these points.
I would like to underscore, very shortly,that the  problem is to 
learn how to  represent  $\cal Z$ at non--zero temperature 
and  baryon density, and how to design a calculational
scheme.

\subsection{ $\cal Z$ at non--zero temperature and chemical potential }

$\cal Z$ is the trace of the density matrix of the system $\hat \rho$
\begin{eqnarray}
{\cal Z} &=& Tr \hat \rho \\ 
\hat \rho &=& e^{(-H - \mu \hat N)}/T
\end{eqnarray}
$H$ is the Hamiltonian, $T$ is the temperature and $\hat N$ is any
conserved number operator. 

\subsection{Chemical potential, relativistic and non-relativistic}

It is worth stressing the main differences 
between the non relativistic and relativistic meaning of
a chemical potential $\mu$, eqs (5) and (8). 

In a non relativistic setting, 
the chemical
potential tells us the energy `cost' of adding an extra particle
to the system: we thus have a different chemical potential
for each type of particle. The term which is added 
to the Hamiltonian is $\mu_k N_k$, $N_k$ being indeed the
number of particles belonging to the $k$--th species.

In a relativistic setting, particles
can be created and destroyed, thus losing their
individuality, so to speak: it only makes sense to have a chemical
potential coupled to the 0-th component of a conserved current.

In practice, we will mostly concern ourselves with the current 
$J_\mu = \bar \psi \gamma_\mu \psi$, 
i.e. $J_0 = \bar \psi \gamma_0 \psi = 
\psi ^\dagger \psi$  will be the density of
 fermion (baryon) number, or,  more precisely, 
the difference between the number of fermions and antifermions:
$\int J_0 = N - \bar N$ .
The chemical potential will then appear in the Lagrangian 
as a term 
$\mu J_0$ and we recognize immediately
that a change of sign for the chemical potential corresponds
to an exchange particles--antiparticles. 

A second difference between non-relativistic and relativistic approach
concerns the magnitude itself of $\mu$, which in a relativistic
theory contains the particle rest mass. 

From a technical point of view, a finite density is easily
handled in a non--relativistic setting. From a relativistic field
theoretic perspectives there are instead important technical problems.
We will discuss these problems later on, and we just anticipate here
that the explicit breaking of the particle--antiparticle symmetry
induced by a chemical potential induces the loss
of positivity of the effective gluonic 
Action--i.e. the breakdown of the condition which makes 
lattice calculations possible.

\subsection{Temperature}

Consider the transition amplitude for returning to the original state
$\phi_a$ after a time $t$:
\begin{equation}
<\phi_a| e^{-iHt}| \phi_a> = \int d\pi
\int_{\phi(x,0) = \phi_a(x)} ^ {\phi(x,t) = \phi_a(x)} d \phi 
e^{i\int_0^tdt \int d^3x (\pi(\vec x, t)
\frac {\partial \phi (\vec x, t)} {\partial t} - H (\pi, \phi))}
\end{equation}
Compare now the above with  expression (2)
for $\cal Z$, and make the trace explicit:
\begin{equation}
{\cal Z} = Tr e^{-\beta ( H - \mu \hat N)} =
\int d \phi_a <\phi_a | e^{-\beta (H - \mu N)} | \phi_a >
\end{equation}
We are naturally lead to the identification
\begin{equation}
\beta \equiv \frac{1}{T} \rightarrow it
\end{equation}

We note - anticipating the discussions of Section below -- 
that studying nonzero temperature on a lattice
\cite{lattice_rev} is straightforward:
one just takes advantage of the finite temporal extent of
the lattice, while keeping
the space directions much larger than any physical scale 
in the system. 

By introducing the integral 
$S(\phi, \psi)$ of the Lagrangian density (from now on we will
always use 1/T as the upper extreme for the time integration)
\begin{equation}
S(\phi, \psi) = \int_0^{1/T} dt \int d^3 x {\cal L}(\phi, \psi)
\end{equation}
$\cal Z$ is  written as
\begin{equation} 
{\cal Z} = \int d \phi d \psi e^{-S(\phi, \psi)}
\end{equation}
The only missing ingredient are the boundary conditions for
the fields: basically, they follow from the (anti)commuting
properties of the (fermi)bose fields
which imply
\begin{equation}
\hat \phi(\vec x, 0) = \hat \phi (\vec x, \beta)
\end{equation}
for the bosons and
\begin{equation}
\hat \psi(\vec x, 0) = -\hat \psi (\vec x, \beta)
\end{equation}
for the fermions.

Fermions and bosons obey antiperiodic and periodic 
boundary conditions, respectively,  in the
time direction.

The expression above, together with the boundary
conditions just introduced, is the key to field 
theory thermodynamics.

\section{Universality}

It is intuitive that when the smallest
significant length scale
of the system $l >> 1/T$ the system becomes effectively 
d--dimensional. Moreover,  the description of the system can be 
effectively `coarse grained', ignoring anything which happens on
a scale smaller than  $l$ .

This can become true when
the system is approaching a continuous transition:
the correlation length of the system $\xi$ is diverging.
In such situation all the physics is dominated by long wavelength
modes. Not only the system gets effectively reduced, but the
coarse graining procedure become doable. As an effect of this
procedure, systems which are very different one from another
might well be described by the same model, provided that the
long range physics is regulated by the same global symmetries: 
this is the idea of universality which provides the 
theoretical framework for
the study of the QCD transition in two interesting (albeit
non physical) cases which we review in the next two sections.

\section{QCD chiral transition ($m_q = 0$)}

Let us recall the symmetries of the QCD action with
$N_f$ flavors of massless quarks, coupled to a $SU(N_c)$ 
color group:
\begin{equation}
SU(N_c)_C \times SU(N_f) \times SU(N_f) \times Z_A(N_f)
\end{equation}
$SU(N_c)$ is the gauge color symmetry.
$SU(N_f) \times SU(N_f) \times Z_A(N_f)$ is
the flavor chiral symmetry, after the breaking of the classical
$U_A(1)$ symmetry to the discrete $Z_A(N_f)$.

We want to study the realization and pattern(s) of breaking
of the chiral symmetries 
and we would like to know the interrelation of the above
with the possibility of quark liberation predicted 
at high temperature and density.

\subsection{Ordinary conditions: zero temperature and density}

In normal conditions (zero temperature and density) the
$SU(N_f)_L\times SU(N_f)_R$ chiral symmetry is spontaneously broken
to the diagonal $SU(N_f)_{L+R}$.Let us 
note the isomorphism:
\begin{equation}
SU(2) \times SU(2) \equiv O(4)
\end{equation}
which shows that the symmetry is the same as the one of an O(4) ferromagnet.
The relevant degrees of freedom are the three pions, and the sigma
particle, and the effective potential is a function of 
$\sigma^2 +  |\pi|^2$ in the chiral space. Once a direction in the
chiral sphere is selected (say in the $\sigma$ direction) chiral
symmetry is spontaneously broken in that direction, according
to the equivalent patterns:
\begin{eqnarray}
 SU(2)_R \times SU(2)_L & \rightarrow & SU(2)_{L + R} \\
 O(4) & \rightarrow & O(3) 
\end{eqnarray}
Massless Goldstone particles (in this case, the three pions) 
appear in the direction orthogonal to the one selected by the
spontaneous breaking.

\subsection{Increasing T}

Disorder increases with temperature. Then, one picture of the high $T$
QCD transition can be drawn by using a ferromagnetic analogy of
the chiral transition: 
$\bar \psi \psi$ can be thought of as a spin
field taking values in real space, 
but whose orientation is in the chiral sphere. 
Chiral symmetry breaking occurs
when $< \bar \psi \psi> \ne 0 $, i.e. it corresponds to the ordered phase.
By increasing T, $< \bar \psi \psi> \to 0$, and the O(4) symmetry
should be restored.

Combining this symmetry analysis with the general idea of dimensional
reduction, Pisarski and Wilczek 
\cite{piwi} proposed that the high temperature transition
in two flavor QCD should be in the universality class of the O(4)
sigma model in three dimensions.
At high temperature when symmetry is restored there will be just one global
minimum for zero value of the fields, and  pion and sigma become eventually 
degenerate.

We have however to keep in mind possible sources of violation of
this appealing scenario and, all in all, 
 one has to resort to numerical simulations to measure
the critical exponents, and verify or disprove the $O(4)$ universality.
In turn, this gives information on the issues raised for instance in
as well as on the possible restoration of the axial
anomaly, see the discussion in Meggiolaro's review.

In practice, one measures the chiral condensate as a function of
the coupling parameter $\beta$, which in turns determines the temperature
of the system. This gives the exponent $\beta_{mag}$ according to
\begin{equation}
<\bar \psi \psi> = B (\beta - \beta_c)^{\beta_{mag}}
\end{equation}
The exponent $\delta$ is extracted from the response at criticality:
\begin{equation}
<\bar \psi \psi> = A m^{1/\delta} ; \beta = \beta_c
\end{equation}
The results  for the critical exponents
compare favorably with the O(4) results
$\beta_{mag} = .38(1)$, $\delta = 4.8(2)$,
and definitively rule out mean fields exponents 
(which would have characterized
a weak first order transition). 
However, the results can still be compatible
with O(2) exponents, which would signal the persistence of some lattice
artifact, and of course it
is still possible that the final answer do not fit any of the above
predictions, for instance if we were just observing some crossover
phenomenon.

In conclusion the symmetry analysis of the (two flavor) QCD transition
gives a definite predictions for the value of the critical exponents,
which is possibly (slightly) violated by the numerical results.
The numerical relevance of this violation (i.e. possible systematic
errors) as well as the its physical implications are an interesting
open problem.

The contribution by E. Meggiolaro covers in more detail these points.

\subsection 
{Increasing density}
Till two or three years ago, we thought that asymptotic freedom 
was the main physical agent behind the pattern of chiral symmetries
at high density. Now it has been recognized that, at least at
zero temperature,  the
main  mechanisms are instability at the Fermi surface leading
to color superconductivity \cite{vari}.
The review by Nardulli and Sannino cover these points which,
unfortunately, are not yet amenable to a lattice study.
At non--zero density the lattice approach
(see contribution by M. D'Elia for developments presented
at this meeting)
is still limited to rather  high temperature, where 
color superconductivity is most likely lost anyway.

\section{QCD deconfinement transition ($m_q = \infty$)}

When $m_q = \infty$  quarks are static and  do not
contribute to the dynamics: hence, the dynamic of the system is 
driven by gluons alone, i.e. we are dealing with a purely Yang-Mills
model:
\begin{equation}
S = F_{\mu \nu}F^{\mu \nu}
\end{equation}
 In addition to the local gauge symmetry, the action enjoys the global 
symmetry associated with the center of the group, $Z(N_c)$. 
The order parameter is the Polyakov loop $P$
\begin{equation}
P = e ^ { i\int_0^{1/T} A_0 dt} 
\end{equation}
In practice, $P$ is the cost of a static source violating the $Z(N_c)$
global symmetry.

The interquark potential V(R,T) (R is the distance, T is the temperature)
is
\begin{equation}
e^{-V(R,T)/T} \propto < P(\vec 0) P^\dagger (\vec R) >
\end{equation}
Confinement can then be read off the behavior of the interquark potential
at large distance.
When  $V(R) \propto \sigma R$  it would cost an infinite
amount of energy to pull two quarks infinitely apart. Above a certain
critical temperature $V(R)$ becomes constant at large distance: i.e.
the string tension is zero, confinement is lost. 
The implication of this is that $|P|^2 = V(\infty, T)$ 
is zero in the confining phase, different from zero otherwise.
P plays then a double role, being the order parameter of the
center symmetry, and an indicator of confinement. 
We learn that in Yang Mills models there is
a natural connection between confinement and realization of the
$Z(N_c)$ symmetry. Hence,  the confinement / deconfinement
transition in Yang Mills systems is amenable to a symmetry
description.
By applying now the same dimensional reduction argument as above,
we conclude that the Universality class expected of
the three color model is the same as the one of a three
dimensional model with $Z(3)$ global symmetry:
Indeed, work by the Columbia, Tsukuba and APE group
in the mid 80's -- see again Nicola Cabibbo's talk
for comments on this -- found that the transition turns 
out to be `almost' second order, i.e. very weakly first order,
like the 3d three state Potts model.

The same reasoning tells us that the two color model is in the universality 
class of the three dimensional $Z(2)$ (Ising) model. 
This prediction has been checked with
a remarkable precision by Engels and collaborators\cite{engels},  
and it is a spectacular confirmation of
the general idea of universality and dimensional reduction.

\section{Summary and Open Questions for the QCD High T Transition}
What do we know about the real world: two (nearly) massless quark
$m_q << \Lambda_{QCD}$, and one more heavy?

We can approach then the 'real' world from two sides, 
either decreasing the mass from infinity, or increasing the
quark mass from zero.

\subsection{Approaching the physical point from infinite mass}
Remember that 
in the infinite mass limit QCD reduces  to the pure gauge
(Yang Mills) model. Yang Mills systems have
a deconfining transition  associated with the realization of the global 
$Z(N_c)$ symmetry.
This places the system in the Ising 3d universality class for two colors,
and makes the transition weakly first order (near second, in fact) for
three colors. General universality arguments are 
perfectly fulfilled by the deconfining transition.

The $Z(N_c)$ symmetry is broken by the kinetic term 
of the action when the quarks are dynamic ($m_q < \infty$) :
this particular symmetry description of deconfinement
only holds for infinite quark mass. 
When light quarks enter the game, the global $Z(N_c)$ 
symmetry observed at infinite mass is lost, and the simple description 
of confinement
in terms of such symmetry is not possible any more. 
It should however remain true that
color forces at large distance should decrease with temperature: the main
mechanism, already at work at $T=0$, is the recombination
of an (heavy) quark and antiquark with pairs generated by the
vacuum: $\bar Q Q \rightarrow \bar q Q + q \bar Q$. 
At high temperature it becomes easier to produce light
$\bar q q$ pairs from the vacuum, 
hence it is easier to `break' the color string  
between an (heavy) quark  and antiquark $\bar Q Q$. 
In other words, we expect enhanced screening of the color forces,
which should be sharp at a  phase transition (or rapid crossover).
It is however worth mentioning that, even if the string `breaks' 
bound states might well survive giving rise to a complicated, 
non--perturbative dynamics above the critical temperature.
The physical scale of these phenomena is  the larger physical
scale in the system, i.e. the pion radius.

\subsection{Approaching the physical point from zero mass}
For zero bare mass the phase transition is chiral.
For three colors, two flavors is second order with $T_c
\simeq 170$ MeV. The prediction from dimensional reduction + universality
--$O(4)$ exponenents-- 
is compatible with the data, but the agreement is not perfect.

If the agreement were confirmed, that would be an argument in favor
of the non-restoration of the $U_A(1)$ symmetry at the transition,
which is also suggested by the behavior of the masses spectrum.
Remember in fact that the chiral partner of the pion is
the $f_0$, which is in turn degenerate with the scalar
$a_0$ with $U_A(1)$ is realized. All in all, 
$U_A(1)$ non--restoration across the chiral transition 
corresponds to $m_\pi \simeq m_{f_0} 
\ne m_{a_0}$ which is the pattern observed in lattice calculations
(once more I refer to meggiolaro's talk).

The transition with three (massless) flavor  turns out to be first
order. The question is than as to whether the strange quark should be 
considered `light' or `heavy'. In general, the real world will be somewhere
in between two and three light flavor, and to really investigate
the nature of the physical phase transition in QCD one should work
as close as possible to the realistic value of the quark masses,
which is a very demanding numerical task.

\subsection{What do we know on the real QCD phase diagram}

Among the most prominent open questions, there is of course the
behavior of `real' QCD, with two light flavor, and a third one
of the order of $\Lambda_{QCD}$, so how and when exactly the 
$N_f=2$ scenario morfs with the $N_f = 3$? Also, why is
$T_\chi$ much smaller that the pure gauge deconfining transition?

At a theoretical level the question is if it is possible to give
an unified description of the two transitions,
chiral and deconfining. This question is
currently under active investigation:
 recent work suggests that a symmetry analysis of the deconfining
transition can be extended also to theories with dynamical fermions.
The physical argument is rooted in a duality transformation
which allows the identification of magnetic monopoles as
agent of deconfinement. The order parameter for deconfimenent
would that be the monopole condensate \cite{deldebbio}.
An alternative approach uses percolation as the common 
agent driving chiral and confining transitions
\cite{satz}.

One unifying description and perhaps the most dramatic evidence
of a phase transition away from the `simple' limits $m_q = 0$
and $m_q = \infty$  comes from the equation of state: for
$T \simeq 180 $ Mev we observe, from lattice calculations,
a sharp increase of the internal energy:
the behavior of the internal energy is a direct probe of the
number of degrees of freedom, and indicates quark and gluon liberation
\cite{lattice_rev}.

Finally, the work \cite{criqcd}
 arrives at an
interesting picture of the phase
diagram of QCD by combining symmetry analysis and phenomenological
consideration. Particularly interesting is the prediction 
of an endpoint of a first order line stemming from zero temperature
chiral transition at finite density, which should be experimentally
observable.

\section{Methods for QCD at finite temperature and density}
Here we will concern ourselves with computational schemes for QCD.
The methods described here are those used to obtain the results
reviewed above on the chiral transition, deconfinement and equation
of state in QCD. We will give some details on these methods, we 
will explain why they are not immediately applicable at finite density and
we will close up with a brief assessment of the current situation
for finite density QCD.

The question is how to estimate the physical observables
$
<O> = Tr O \hat \rho / {\cal Z} \nonumber
$
starting from the representation of the partition function

\begin{equation}
{\cal Z}(\mu, T) = \int_{0}^{1/T} dt \int e^{
 -S_G + 
\bar \psi (\not \partial + m + \mu \gamma_0) \psi}
d \bar \psi d \psi dU \nonumber
\end{equation}

The need for  two integrations (over bosons and over
fermions) lends itself naturally to two different paths:
either integrate gluons first, or fermions.

In the first case (if gluons are integrated out first) 
we end up with a purely effective fermionic model:
\begin{equation}
{\cal Z} (T, \mu, \bar \psi, \psi, U) \simeq {\cal Z} (T, \mu, \bar \psi, \psi)
\end{equation}
As the integration over gluons 
cannot be done exactly 
\footnote{ There indeed a a systematic
approach to QCD based on the strong coupling expansion which makes
this integrations easy, but this is not in the scope of 
this introduction},
 such effective models are often build with the help of a symmetries'
analysis for QCD. The gauge fields enters the the game
under the guise of coefficients for such models. 
Very important examples
of this approach include instanton models, chiral perturbation
theory, four fermion models
such as those discussed by Nardulli and Sannino at this
meeting.

The other way takes advantage of the bilinear nature 
of S in the fermionic fields, yielding the exact expression:
\begin{equation}
{\cal Z}(T, \mu, U) = 
\int dU e ^{-(S_g - log(det M))}
\end{equation}
The above integral needs being regulated, and 
scheme for calculating 
it has to be devised. Both tasks are accomplished within the
lattice approach.

\subsection{ Lattice field theory at T,$\mu \ne$ 0}

Temperature comes for free on a lattice: the lattice has a
finite extent $N_t a$, hence temperature is given by
$T = 1 / N_t a $. The discretization can be carried
out in complete analogy to $T=0$, and most of the techniques
developed there (see e.g. the recent review 
for an introduction to lattice field theory at zero temperature) 
apply at finite temperature as well.

A finite density of baryons $\mu_B J_0$ 
\footnote{
Remember again that $L(\mu) = L_0 + \mu J_0$,
$J_0 = \bar \psi \gamma_0 \psi$, i.e. $ N - \bar N = \int J_0$}
poses instead specific problems. 
\vskip .5 truecm
\noindent
\underline{How to discretize $\mu \bar \psi \gamma_0 \psi$}

Recall first the `natural' discretization of the matter fields
$\phi(x)$  and their derivatives $\partial_\mu \phi(x)$ 
on a regular lattice with spacing $a$:

\begin{eqnarray}
\phi_{LATT}(n_1, n_2, n_3, n_4)& = &\phi(n_1a, n_2a, n_3a, n_4a)  \nonumber \\
\Delta_\mu
\phi_{LATT}(n_1, n_2, n_3, n_4) & = &  
 (\phi(n_1a, (n_\mu + 1)a, n_3a, n_4a) - \nonumber \\ && 
\phi(n_1a, (n_\mu a, n_3a, n_4a))/2a \nonumber
\end{eqnarray}

This, for instance, is the correct 
prescription for the chiral condensate:
$ m \bar \psi \psi \rightarrow m \bar \psi_{LATT} \psi_{LATT}$
but not, as we will see in a moment, for the baryon density!
$ \mu \bar \psi \gamma_0 \psi \rightarrow \mu \bar \psi_{LATT} \gamma_0
\psi_{LATT} $ is not the correct lattice form , i.e.
{\em The naive discretization is not adequate for baryon density}.

Let us then consider free fermions in the \underline{continuum} 
$$S = \int_0^\beta \bar \psi \gamma_mu \partial_\mu \psi
+ m \bar \psi \psi + \mu \bar \psi \gamma_0 \psi$$

The internal energy 
$\epsilon = \frac {1}{V} \frac {\partial}{\partial \beta} ln 
{ \cal Z} 
= \frac {4}{(2 \pi) ^4 } \int d^4 p 
 \frac 
{ (p_0 + i \mu)^2 } 
{ (p_0 + i \mu)^2 + p^2 + m^2 }$,
after subracting the vacuum energy, is finite at 
$T=0$: 
$ lim _{T \to 0} \epsilon =  \frac {\mu^4}{4 \pi^2}$
and gives the expect result

By use of the naive discretization,
$\epsilon$ 
would instead diverge in the continuum ($a \to 0$):
$ L = \bar \psi_x  
\gamma_\mu \psi_{x + \mu a} + m \bar \psi_x \psi_x +
 \mu \bar \psi_x \gamma_0 \psi_x \epsilon  
\propto \frac{\mu^2}{a^2} \to_{a \to 0} \infty$

\vskip .4 truecm
\noindent
\underline { The solution \cite{lattice_rev}: $\mu$ is an external field} 

Note the analogy:
$\bar \psi \gamma_\mu A_\mu \psi \leftarrow \rightarrow  
i \mu \bar \psi \gamma_0 \psi$. It shows us that
$\mu$ looks like an external field in the time direction.
But we know how to 'put' external field on a lattice: they
live on the lattice links: for instance, in electrodynamics 
$A  \to \theta = e^{ i A} $  (and, for the gauge fields, this
implements in a natural and elegant way gauge invariance)

These considerations suggest how to put finite density on a lattice:
\begin{equation} 
L(\mu) = \bar \psi_x \gamma_0 e^{\mu a}\psi_{x + \hat 0} -
      \bar \psi_{x + \hat 0} \gamma_0 e^{-\mu a}\psi_{x }
\end{equation} 
Indeed, this turns out to be the correct prescription: unphysical
divergences disappear and the  continuum limit is reproduced.

There is also a very expressive physical interpretation: as we can see,
forward propagation is encouraged and backward propagation is suppressed:
we are indeed inducing an asymmetry particles-antiparticles.

The 0-th component of the current $j_0$ counts indeed the differences
between backward and forward propagation:
\begin{equation}
J_0 = - \partial_\mu L = \bar \psi_x \gamma_0 e^{\mu a}\psi_{x + \hat 0} +
      \bar \psi_{x + \hat 0} \gamma_0 e^{-\mu a}\psi_{x }
\end{equation}
and reproduces the correct continuum limit.

\vskip .4 truecm
\noindent
\underline{ Chemical potential and boundary conditions}

By use of an unitary transformation of the fields it is possible
to re-express $L$ as

\begin{equation}
 L (\mu) = L (0) 
\end{equation}
with boundary conditions
\begin{eqnarray}
\phi(x + N_T) &=& e^{\mu N_T} \phi(x) \\
\psi(x + N_T) & = & -e^{\mu N_T} \psi(x)
\end{eqnarray}

It is of some interest to consider the effect
of the chemical potential  on the baryonic propagators in the
two cases -- when the chemical potential is included
into the Action, and when instead affects only the boundary
conditions. 
In the first case, 
(at zero temperature, and ignoring feedbacks) a term
$exp(-N_c \mu)$ multiplies the baryonic propagator: 
this produces an apparent decrease of the baryon mass
\begin{equation}
m_B = m_B - 3 \mu
\end{equation}
i.e. the baryon becomes massless at $\mu = \mu_c = m_B/3$. 
In the same situation, the chiral condensate remains constant till 
$\mu_c$, and then suddenly drops to zero.
The behavior of the chiral condensate would be
suggestive of a strong first order transition, while the behavior of
the baryon mass would suggest a second order transition with $\nu = 1$!

In the other formulation
(when the chemical potential only affects the boundary conditions), 
at zero temperature the baryon mass is constant till $\mu_c$ : 
this is consistent with the behavior of the chiral condensate,
and with the physical intuition that nothing should happen till
the Fermi level is reached. 
Of course, nothing is wrong and the apparent contradiction is
resolved by noticing that the apparent decrease of the baryon mass
below $\mu_c$ merely reflects the change of the reference energy.
The relationships
above provide a link between the two pictures.

\vskip .4 truecm
\noindent
\underline{ QCD at nonzero T and $\mu$ at a glance}

The continuum formulation:
\begin{eqnarray}
\nonumber
 L &=& L_{YM} +  \bar \psi  (i \gamma_\mu D_\mu + m) \psi \nonumber \\
 \nonumber  & + & \mu \bar \psi \gamma_0 \psi \nonumber \\
\end{eqnarray}
$\mu$ is explicitly included via the coupling to $\mu J_0$ , and
the temperature is the reciprocal of the imaginary time.

On a lattice:
\begin{eqnarray}
L & = &  L_{YM} + \sum_{i = 1}^3 \bar \psi_x U \gamma_i \psi_{x + \hat i} -
      \bar \psi_{x + \hat i} U^\dagger \gamma_i \psi_{x }  
\nonumber \\ & + &  m \bar\psi\psi +
\bar \psi_x \gamma_0 e^{\mu a} U \psi_{x + \hat 0} -
      \bar \psi_{x + \hat 0} U^\dagger \gamma_0 e^{-\mu a}\psi_{x } \nonumber
\end{eqnarray}

$\mu$ appears as a link term, and the temperature is again the reciprocal
of the imaginary time. The two formulations coincide in the limit
$a \to 0$.

\subsection {From the formulation to the results }

Let us write again:
\begin{equation}
S = S_{YM}(U) + \bar \psi M (U) \psi
\end{equation}
By taking advantage of $S$ bilinearity in the fields $\psi, \bar \psi$
we can write
\begin{equation}
{ \cal Z } = \int e^{-S_{YM}(U)} det M(U) dU 
\end{equation}
It is convenient to define an 'effective' Action
\begin{equation}
S_{eff}(U)= S_{YM}(U) - ln( det M(U)) 
\end{equation}
Averages of purely gluonic observables can be expressed as
\begin{equation}
<f(U)> = Z^{-1} \int dU e^{-S_{eff}(U)} f(U) 
\end{equation}
while fermion bilinears can be evaluated with the help
of Grassman algebra
\begin{equation}
<\bar q q > = Z^{-1} \int dU e^{-S_{eff}(U)} Det M(U)^{-1} 
\end{equation}
In conclusion, if we know how to treat $\int e^{-S_{eff}(U)}  dU $ 
we  have access to all of the gluonic observables, chiral condensate,
meson and baryons propagators (masses, decay constant, etc.) etc.
much in the same way as at $T=0$.

In practical numerical works lattice discretization is combined with
importance sampling: a configuration of gauge fields [U]
 is a 'point'
in a multidimensional integration space. A Markov chain of points
is then created according to the prescription:
    $P([U]) \propto e ^{-S_{eff}([U])}$
Expectation values are then given by simple averages:
$<O> = lim_{N \to \infty} 1/N \sum_{i = 1}^N O(U) $.

The prescription above relies on importance sampling: $S_eff([U])$
must be positive \footnote{If this is not the case, one might think of using
$S^\dagger$ instead: this has to be used with care because
it introduces extra degrees of freedom which might well be dangerous.}
In QCD (eq. above)  $M^ \dagger (\mu) = - M  (-\mu)$ : 
importance sampling, hence MonteCarlo evaluation of physical observables,
works for purely imaginary chemical potential, including, of course, 
$\mu=0$.

For QCD with two color, however, as well as for several fermionic models,
the positivity condition is met by the Action at non zero $\mu$. All in
all, we have a rather elaborate pattern of warnings and possibilities.

\subsection {Results and possibilities for QCD at $\mu \ne 0$ }

I would classify the many questions one can ask along two main lines:

Firstly, what can be done in order to learn about the{\em general} behavior at 
nonzero baryon density?

Secondly,  what can be done in order to learn about  {\em real} QCD at 
nonzero baryon density?

I would say that to learn 
about the effects of a finite density of baryons, we can study
two color QCD : here, not only one can study fermionic observables, but it is
also possible to observe the effects of a density of baryons on the gauge 
fields. This is indeed an active field of research, and I refer to the
proceedings of the Lattice meetings for details.

How about {\em real} QCD? there is a growing consensus that at least the high
temperature, small density region is accessible: by increasing temperature
it is easier to fluctuate light baryons, hence to explore a region with
a moderate baryon density. At
least three different, complementary approaches stem
from this consideration:direct evaluation 
of derivatives at $\mu$ = 0, reweighting, analytic continuation from
purely imaginary chemical potential~\cite{kogut}. 
This last approach has been discussed
by M. D'Elia at this meeting. 

\vskip .2 truecm
I hope to have managed at least to describe some of the
aspects of a rapidly evolving
field, and to convey a feeling of its many reasons of interest.
A discussion of the recent  results is not in the
scope of this cursory introduction: for this, and for a complete
set of references, 
I refer to the latest reviews \cite{kogut} (not yet available at the time
of the Frascati meeting).


\begin{thebibliography}{99}

\bibitem
{cabibbo} N. Cabibbo, introductory remarks, this volume.  
\bibitem{satz} H. Satz, ibid.
\bibitem{nardulli} G. Nardulli, ibid.; M. Mannarelli, ibid.  
\bibitem{sannino} F. Sannino, ibid.; W. Schaefer, ibid.  
\bibitem{deldebbio} L. Del Debbio, ibid.; A. Drago, ibid.; D. Cosmai,
ibid. ; A. Papa, ibid.  
\bibitem{meggiolaro} E. Meggiolaro, ibid.  
\bibitem{delia} M. d'Elia, ibid.  
\bibitem{kogrev} Introductory material on the lattice approach
to field theory can be found in
J. B. Kogut, Rev.Mod.Phys.51 (1979) 659;  M. Creutz, 
{\em Quarks, Gluons and Lattice}, Cambridge;
C. Itzykson-Douffre, {\em Statistical Field Theory},
Cambridge,  Chapter 7.
\bibitem{vari} M. Alford, 
{\tt hep-ph/0102047};  K. Rajagopal 
and F. Wilczek, 
in 'At the Frontier of Particle Physics/Handbook of QCD', 
M. Shifman, ed., (World Scientific); 
R. Rapp, T. Schafer, E.V. Shuryak, M. Velkovsky,
 Annals Phys.280 (2000)35.
\bibitem{lattice_rev} For an
 introduction to numerical simulations
of lattice QCD thermodynamics see e.g.
C. DeTar, in Hwa, R.C. (ed.): 
Quark-gluon plasma, vol.2 , pag. 1 or
F. Karsch, hep-lat/0206034.
\bibitem{piwi} 
R. D. Pisarski and F. Wilczek , Phys.Rev. D29 (1984)338.
\bibitem{engels} 
J. Engels, S. Mashkevich, T. Scheideler, G. Zinovev,
Phys.Lett.B365 (1996) 219.
\bibitem{criqcd} M. Stephanov, K. Rajagopal, E. Shuryak, Phys. Rev. Lett
Phys.Rev.Lett.81 (1998) 4816; 
J. Berges and K. Rajagopal, 
Nucl.Phys. B538 (1999) 215;
 	M.A. Halasz, A.D. Jackson, R.E. Shrock, M.A. Stephanov 
 	and J.J.M. Verbaarschot, Phys.Rev.D58
(1998) 096007; M. Stephanov, K. Rajagopal, E. Shuryak, 
Phys.Rev.D60 (1999) 114028.
\bibitem{kogut} Recent reviews on QCD in exteme conditions include:
J. B. Kogut, Lattice2002; Z. Fodor, QM2002; K. Kanaya, QM2002;
M. Alford, ICHEP2002.
\end{thebibliography}
\end{document}